\documentstyle[aps,preprint]{revtex}

\begin{document}
\draft \tightenlines \preprint{gr-qc/0005019}

\title{Classical and Quantum Surgery of Geometries in an Open Inflationary Universe}
\author{Sang Pyo Kim\footnote{Electronic address: sangkim@ks.kunsan.ac.kr}}
\address{Department of Physics \\ Kunsan National University
\\ Kunsan 573-701, Korea}

\date{\today}

\maketitle
\begin{abstract}
We study classically and quantum mechanically the Euclidean
geometries compatible with an open inflationary universe of a
Lorentzian geometry. The Lorentzian geometry of the open universe
with an ordinary matter state matches either an open or a closed
Euclidean geometry at the cosmological singularity. With an exotic
matter state it matches only the open Euclidean geometry and
describes a genuine instanton regular at the boundary of a finite
radius. The wave functions are found that describe the quantum
creation of the open inflationary universe.
\end{abstract}
\pacs{PACS number(s): 98.80.H, 98.80.Cq, 04.60Ds, 04.60Kz}

Current data from large-scale structure and the cosmic microwave
background suggest open inflationary universe models as a viable
theory \cite{silk}. Recently Hawking and Turok (HT) proposed such
an open inflationary model by using a singular gravitational
instanton \cite{hawking-turok}. This raised, however, a hot debate
whether the HT instantons can lead to the most probable wave
functions in quantum cosmology
\cite{linde,hawking-turok2,vilenkin}. Vilenkin argued that the
HT-type singular instantons of a massless scalar model may lead to
a physically unacceptable consequence \cite{vilenkin2}. The key
issue of the debate is whether one may find the instantons of a
Euclidean geometry that are well behaved and match the open
inflationary universe of a Lorentzian geometry. Hence, the
condition necessary for the surgery of Euclidean and Lorentzian
geometries is essential in understanding properly the open
inflationary models.

In this Letter we study the classical and quantum compatibility
condition in gluing Lorentzian and Euclidean geometries. Our model
is an open FRW universe of the Lorentzian geometry with a minimal
massless scalar field and a cosmological constant. The genuine
quantum creation of the Universe requires the backward evolution
either to the cosmological singularity or to a gravitational
potential barrier at a finite size, in which it fits with a
Euclidean universe of the same size. The HT instanton solutions
\cite{hawking-turok,hawking-turok2} and the asymptotically flat
solution by Vilenkin \cite{vilenkin2} belong to the former case.
The latter case will also be treated in this paper.

Let us begin with an open FRW universe described by the action
\begin{equation}
I =  \frac{m_P^2}{16 \pi} \int d^4x \sqrt{-g}  \Bigl[R - 2 \Lambda
\Bigr] + \frac{m_P^2}{8 \pi} \int d^3x \sqrt{h} K -
\frac{1}{2}\int d^4x \sqrt{-g} g^{\mu \nu}
\partial_{\mu} \phi \partial_{\nu} \phi, \label{action}
\end{equation}
where $m_P = 1/\sqrt{G}$ is the Planck mass, and $\Lambda$ and
$\phi$ denote the cosmological constant and the scalar field,
respectively. The sign in Eq. (\ref{action}) has been chosen to
yield a positive definite energy density for the scalar in
Lorentzian geometries, and the surface term for the gravity has
been introduced to yield the correct equation of motion for the
closed universe.

The open Lorentzian FRW universe has the metric
\begin{equation}
ds_L^2 = - N^2 (t) dt^2 + a^2 (t) \bigl[d\xi^2 + \sinh^2\xi d
\Omega^2_2 \bigr], \label{lor met}
\end{equation}
where $N(t)$ is the lapse function, $d\Omega^2_2$ is the standard
metric on the unit two-sphere $S^2$, and $\xi$ ranges over $(0,
\infty)$. Now the action (\ref{action}) becomes
\begin{equation}
I_{L} = \int dt \Bigl[d\xi \frac{2}{\pi} \sinh^2 \xi \Bigr]
\Biggl[- \frac{3 \pi m_P^2}{4} \frac{a \dot{a}^2}{N} - \frac{3\pi
m_P^2}{4} N \Biggl(a - \frac{\Lambda}{3}a^3 \Biggr) + \pi^2
\frac{a^3 \dot{\phi}^2}{N} \Biggr], \label{lo act}
\end{equation}
where dots denote derivative with respect to $t$. Here the volume
$\int d\xi [2\sinh^2 \xi/\pi]$ is factored out to match an open
geometry with a closed one\footnote{The additional factor $2/\pi$
is factored out to yield the same kinetic terms for the gravity
and the scalar field in the Wheeler-DeWitt equation for the open
and the closed topology. But this does not change the classical
equations of motion and even the semiclassical equations from the
Wheeler-DeWitt equation.}, and the total derivative term is
canceled by the surface term. By varying the action (\ref{lo act})
with respect to $N$ and $\phi$ one obtains the time-time component
of Einstein equation and the scalar field equation
\begin{eqnarray}
\Biggl(\frac{\dot{a}}{a} \Biggr)^2 - \Biggl(\frac{\Lambda}{3} a^2
+ \frac{1}{a^2}\Biggr) = \frac{4 \pi}{3 m_P^2} \dot{\phi}^2,
\label{lo eq} \\ \frac{\partial}{\partial t} (a^3 \dot{\phi}) = 0.
\label{sc eq}
\end{eqnarray}

On the other hand, the Euclidean FRW geometry has the metric
\begin{equation}
ds_E^2  =  N^2 (\tau) d\tau^2 + b^2 (\tau) \bigl[ d\xi^2 +
f^2_{\pm} (\xi) \xi d \Omega_2^2],
\end{equation}
where $ f_+ (\xi) = \sinh \xi$ for an open universe and $f_- (\xi)
= \sin \xi$ for a closed universe. The open (closed) universe
leads to the action (\ref{action})
\begin{equation}
I_{E} = \int dt \Bigl[ d\xi \frac{2}{\pi} \sinh^2 \xi \Bigr]
\Biggl[ \frac{3 \pi m_P^2}{4} \frac{b (b')^2}{N} - \frac{3\pi
m_P^2}{4} N \Biggl(\frac{\Lambda}{3}b^3  \pm  b\Biggr) - \pi^2
\frac{b^3 (\phi')^2}{N} \Biggr], \label{eu act}
\end{equation}
where the volume factor $\int d\xi [2\sinh^2 \xi/\pi]$ is only for
the open universe and primes denote derivative with respect to
$\tau$ and the upper (lower) sign for the open (closed) universe,
respectively. The time-time component of Einstein equation for the
open (closed) universe and the scalar field equation follows from
the actions (\ref{eu act}):
\begin{eqnarray}
\Biggl(\frac{b'}{b} \Biggr)^2 + \Biggl(\frac{\Lambda}{3} b^2 \pm
\frac{1}{b^2} \Biggr) = \frac{4 \pi}{3 m_P^2} (\phi')^2, \label{eu
eq} \\ \frac{\partial}{\partial \tau} (b^3 \phi') = 0. \label{e-sc
eq}
\end{eqnarray}

In the Euclidean geometry, the classical equation of motion
(\ref{eu eq}) for $b$ is obtained by substituting the solution to
the scalar field equation (\ref{e-sc eq})
\begin{equation}
b^3 \phi' = \frac{p}{2 \pi^2}. \label{mom}
\end{equation}
There is a classically allowed motion due to the scalar field for
the open universe and due to the scalar field and the scalar
curvature for the closed universe. However, the classical motion
is limited up to a turning point $b_*$ by the cosmological
constant, which acts as a source for negative energy in the
Euclidean geometry. Therefore, there is a periodic motion  between
the zero and the finite radius. Similarly, in the Lorentzian
geometry, the scalar field equation (\ref{sc eq}) may have two
types of solutions: an ordinary or an exotic state solution
\begin{equation}
a^3 \dot{\phi} = \frac{p}{2 \pi^2}~~ {\rm or}~~ a^3 \dot{\phi} = i
\frac{\kappa}{2 \pi^2}, \label{mat st}
\end{equation}
where $p$ and $\kappa$ are real constants. The imaginary quantity
is the result of the Wick-rotation $t = i \tau$ of Eq. (\ref{mom})
from the Euclidean to Lorentzian geometry and, in fact,
corresponds to a particular quantum state of the scalar field in
quantum cosmology \cite{kim}. In the former case of ordinary
matter states, the classical motion (\ref{lo eq}) extends over
$(0, \infty)$. Hence, the surgery of the Lorentzian geometry with
the Euclidean one should occur at the cosmological singularity $(a
= b = 0)$. On the other hand, in the latter case of exotic matter
states, the classical equation of motion becomes
\begin{equation}
\Biggl(\frac{\dot{a}}{a} \Biggr)^2 - \Biggl(\frac{\Lambda}{3} a^2
+ \frac{1}{a^2}\Biggr) + \frac{\kappa^2}{3 \pi^3 m_P^2}
\frac{1}{a^4} = 0. \label{lo eq2}
\end{equation}
There is a classical turning point $a_*$, and the classical motion
extends over $(a_*, \infty)$. The regime $(0, a_*)$ of the
classically forbidden motion for the Lorentzian geometry now
should be matched with the same regime of the classically allowed
motion (\ref{eu eq}) obtained by substituting Eq. (\ref{mom}) with
$p = \kappa$ for the open Euclidean geometry. This surgery at the
finite radius does not hold for the closed Euclidean geometry, the
turning point of which differs from $a_*$.

At the classical level, the requirement for matching two
geometries is that the radii and the second fundamental form be
continuous across the boundary. First, we consider the surgery at
$b  = a = 0$. For the FRW geometry, the second fundamental form is
given by $K_{ij} = - (a \dot{a}/N) g_{ij}$, the nonvanishing
components of which are $g_{ii} = (1, f_{\pm}^2, f_{\pm}^2 \sin^2
\theta)$. This implies that the the second fundamental forms of
the Euclidean and Lorentzian geometry vanish at the cosmological
singularity. Therefore, the open universe of Lorentzian geometry
can match both the open and the closed universe of Euclidean
geometry that has the boundary $b = 0$ as a turning point of
periodic motion. For instance, the HT-type instanton solution $b
(\tau) = \sqrt{(3/\Lambda)} \cos\sqrt{(\Lambda/3)\tau}$, which is
the limiting case of $p = 0$, i.e. without the scalar field,
starts and ends at $b = 0$. The apparent drawback of the surgery
at $a = b = 0$ is that the scalar field diverges as shown in Eqs.
(\ref{mom}) and (\ref{mat st}), as one approaches to this
boundary, though the total action is finite. The Vilenkin's
asymptotically flat instantons with the scalar field but without
$\Lambda$ are thus singular and may not be physically acceptable
\cite{vilenkin2}. Though Hawking and Turok considered only the
closed Euclidean geometry, the open Euclidean geometry can also be
allowed. The exclusion of the open Euclidean geometry should rest
on another reason such as the infinite volume, but not merely the
surgery itself.

Second, we consider the surgery at a finite radius. When one
matches an open Lorentzian geometry with a closed Euclidean
geometry, there is a discontinuity of the second fundamental form
at the boundary of a finite radius. Moreover, the classical
turning point $b_*$ of the Euclidean geometry differs from $a_*$
of the Lorentzian geometry unless $\Lambda = 0$ and $p = \kappa$.
But even in this case one has $a_* = b_* = 0$. Hence, the open
Lorentzian geometry can not match the closed Euclidean geometry at
a finite radius. However, Eq. (\ref{eu eq}) for the open Euclidean
geometry is the instanton equation for the Lorentzian geometry
(\ref{lo eq2}) with an exotic matter state. Thus, the classically
forbidden regime of $a$ becomes exactly the classically allowed
regime of $b$ and vice versa. For $p = \kappa$, the classical
turning points $a_*$ and $b_*$ are equal and approximately given
by $a_* = b_* \approx [\kappa^2/(3 \pi^3 m_P^2)]^{1/4}$ .
Therefore, at the boundary of a finite radius the open Lorentzian
geometry can only match the open Euclidean geometry. Near the
turning points, the solutions to Eqs. (\ref{lo eq2}) and (\ref{eu
eq}) are approximately given by
\begin{equation}
a (t) \approx a_* + \frac{16}{a_*} (t - t_*)^2, \quad b (\tau)
\approx b_* - \frac{16}{b_*} (\tau_* - \tau)^2,
\end{equation}
where $t_*$ and $\tau_*$ are the Lorentzian and the Euclidean time
at the boundary $a_* = b_*$. One prominent feature is that all
geometric quantities and the scalar field are regular at the
matching boundary.

At the quantum level, the matching condition is the continuity of
the wave function and its first derivative at the boundary, which
is the outcome of the continuity across the boundary of the radius
and the conjugate momentum $\pi_a = - (3 \pi m_P^2a
\dot{a})/(2N)$, in proportional to the second fundamental form.
$N$ being the lapse function, the Lorentzian action (\ref{lo act})
leads to the Hamiltonian density constraint
\begin{equation}
{\cal H}_L = - \frac{1}{3 \pi m_P^2 a} \pi_a^2 + \frac{3 \pi
 m_P^2}{4} \Bigl(a + \frac{\Lambda}{3} a^3\Bigr) +
\frac{1}{4 \pi^2 a^3} \pi_{\phi}^2 = 0, \label{lo ham}
\end{equation}
where $ \pi_{\phi} = \pi^2 a^3 \dot{\phi}/N$. According to the
Dirac quantization procedure, Eq. (\ref{lo ham}) becomes the
Wheeler-DeWitt equation
\begin{equation}
\Biggl[- \frac{\hbar^2}{2m_P^2} \frac{1}{a^{\nu}}
\frac{\partial}{\partial a} \Biggl(a^{\nu}
\frac{\partial}{\partial a}\Biggr)- \frac{9 \pi^2 m_P^2}{8}
\Biggl(\frac{\Lambda}{3} a^4 + a^2 \Biggr) + \frac{3 \hbar^2}{8
\pi a^2} \frac{\partial^2}{\partial \phi^2} \Biggr] \Psi_L (a,
\phi) = 0,
\end{equation}
where $\nu$ denotes some part of operator ordering ambiguity. The
classical exotic state of the scalar field in the Lorentzian
geometry is described by the wave function
\begin{equation}
\Phi^{\epsilon}_{L} = \lim_{\epsilon \rightarrow 0_+} \frac{1}{(2
\pi)^{3/2}} \exp \Bigl[- \kappa \tanh (\frac{\phi}{\epsilon})
\phi/ \hbar \Bigr]. \label{lo ex}
\end{equation}
It has a negative energy density, and bounded having the
asymptotic form $e^{- \kappa |\phi|/\hbar}$ as $\phi \rightarrow
\pm \infty$. The gravitational field equation of the wave function
$\Psi_L = \psi_L (a) \Phi_L (\phi)$ takes the form
\begin{equation}
\Biggl[- \frac{\hbar^2}{2m_P^2} \frac{1}{a^{\nu}}
\frac{\partial}{\partial a} \Biggl(a^{\nu}
\frac{\partial}{\partial a}\Biggr)- \frac{9 \pi^2 m_P^2}{8}
\Biggl(\frac{\Lambda}{3} a^4 + a^2 \Biggr) + \frac{3 \kappa^2}{8
\pi a^2} \Biggr] \psi_L (a) = 0. \label{wd gr1}
\end{equation}
The turning point for Eq. (\ref{wd gr1}) is given by the same
$a_*$. In the region $a \ll \sqrt{3/\Lambda}$ where
$(\Lambda/3)a^2$ is small compared with $a^2$, the wave functions
of Eq. (\ref{wd gr1}) are approximately given by \cite{g&r}
\begin{equation}
\psi_L (a) = a^{(1 - \nu)/2} Z_{\alpha} (\beta a^2),
\end{equation}
where $Z$ are Bessel functions and $ \alpha = [(1 - \nu)^2/4 + (3
m_P^2 \kappa^2)/(4 \pi)]^{1/2}/2$ and $\beta = 3 \pi m_P^2/(4
\hbar)$. The wave function holds for all range $(0, \infty)$. The
Hankel function $ H^{(1)}_{\alpha} (\beta a^2)$ provides an
expanding wave function, which corresponds to the Vilenkin's
tunneling wave function \cite{vilenkin3}. In the tunneling regime
where $\alpha \gg \beta a^2$ or $a \ll a_*$, the tunneling wave
function has an asymptotic expansion by the index
\begin{equation}
\psi^{\rm T}_L (a) \approx -i a^{(1-\nu)/2}\frac{e^{\alpha \gamma
- \alpha \tanh \gamma}}{\sqrt{\pi \alpha \tanh \gamma/2}},
\end{equation}
where $\cosh \gamma = \alpha/(\beta a^2) = a^2_*/a^2$ \cite{g&r}.
On the other hand, the Hartle-Hawking's no-boundary wave function
\cite{hartle-hawking} is prescribed by the Bessel function
$J_{\alpha} (\beta a^2)$, which is a superposition of an expanding
branch $H^{(1)}_{\alpha} (\beta a^2) $ and a recollapsing branch
$H^{(2)}_{\alpha} (\beta a^2) $. For $a \ll a_*$, it is
approximately given by
\begin{equation}
\psi^{\rm HH}_L (a) \approx a^{(1-\nu)/2} \frac{e^{-( \alpha
\gamma - \alpha \tanh \gamma)}}{\sqrt{2 \pi \alpha \tanh \gamma}},
\end{equation}
and is regular at the cosmological singularity.

Finally we turn to matching the wave functions at $a = b = 0$. In
the Lorentzian geometry, the ordinary state of the scalar field is
given by the wave function
\begin{equation}
\Phi_L (\phi) = \frac{1}{(2\pi)^{3/2}} e^{i p\phi/\hbar}.
\end{equation}
Then the gravitational field equation separates as
\begin{equation}
\Biggl[- \frac{\hbar^2}{2m_P^2} \frac{1}{a^{\nu}}
\frac{\partial}{\partial a} \Biggl(a^{\nu}
\frac{\partial}{\partial a}\Biggr)- \frac{9 \pi^2 m_P^2}{8}
\Biggl(\frac{\Lambda}{3} a^4 + a^2 \Biggr) - \frac{3 p^2}{8 \pi
a^2} \Biggr] \psi_L (a) = 0. \label{wd gr2}
\end{equation}
There is no classical forbidden regime and the wave functions are
defined for $(0, \infty)$. The wave functions in the region $a \ll
\sqrt{3/\Lambda}$ were found \cite{kim2}
\begin{equation}
\psi_E (b) = b^{(1 - \nu)/2} Z_{\tilde{\alpha}} (i \beta b^2),
\label{wave3}
\end{equation}
where $\tilde{\alpha}$ is obtained from $\alpha$ by continuing
analytically $\kappa = i p$. For the covariant operator ordering
$\nu = 1$ or large $p$, the index becomes pure imaginary,
$\tilde{\alpha} = i \alpha$. The expanding and the recollapsing
wave function are given by $H^{(1)}_{\tilde{\alpha}} (i \beta
a^2)$ and $H^{(2)}_{\tilde{\alpha}} (i \beta a^2)$. These wave
functions can be smoothly matched with the asymptotic wave
functions in the region $a \gg \sqrt{3/\Lambda}$ in Ref.
\cite{kim3}. On the other hand, in the Euclidean geometry, the
ordinary quantum state $\Phi_E (\phi) =
 e^{i p\phi/\hbar}/(2\pi)^{3/2}$ leads to the gravitational field equation
\begin{equation}
\Biggl[- \frac{\hbar^2}{2m_P^2} \frac{1}{b^{\nu}}
\frac{\partial}{\partial b} \Biggl(b^{\nu}
\frac{\partial}{\partial b}\Biggr) + \frac{9 \pi^2 m_P^2}{8}
\Biggl(\frac{\Lambda}{3} b^4 \mp b^2 \Biggr) - \frac{3 p^2}{8 \pi
b^2} \Biggr] \psi_E (b) = 0, \label{wd gr3}
\end{equation}
where the upper (lower) sign is for the closed (open) geometry,
respectively. Far away from the matching boundary $a = b = 0$, the
cosmological constant term prevails over the other terms and
provides a potential barrier. The wave functions exhibits
exponential behavior at large $b$. But not far away from the
matching boundary $a = b = 0$, the cosmological constant term can
be neglected compared with the other two terms. So, Eq. (\ref{wd
gr3}) for the closed Euclidean geometry is approximately equal to
Eq. (\ref{wd gr2}) for the open Lorentzian geometry and their wave
functions are given by Eq. (\ref{wave3}). Very close to the
boundary, even the curvature terms $a^2$ and $b^2$ can also be
neglected, and Eqs. (\ref{wd gr2}) and (\ref{wd gr3}) are
dominated by the scalar term and are approximately equal to each
other. This means that the open Lorentzian geometry can match
either the closed or the open Euclidean geometry. This is true
also for a general scalar field since the kinetic term dominates
over the potential term and the scalar field becomes roughly
massless and stiff near the cosmological singularity. Therefore,
the tunneling wave function can be matched very accurately at $a =
b = 0$ and the Hartle-Hawking wave function, which vanishes at the
boundary, can be matched exactly. This is the very reason how the
HT instantons of the closed Euclidean geometry fit with the open
inflationary universe of Lorentzian geometry. But another
possibility still remains that was not considered in Refs.
\cite{hawking-turok,hawking-turok2,vilenkin2}: at the finite
radius the open universe with an exotic state can match exactly
the open Euclidean geometry with an ordinary state, which is
nothing but the Wick-rotation of time \cite{kim}. This is exactly
the counterpart of a closed universe of the Lorentzian geometry
matched with the closed Euclidean geometry \cite{kim4}.

In summary, we have studied the classical and quantum matching
condition of Euclidean and Lorentzian geometries. The matching
boundary depends crucially on the states of the scalar field. The
open inflationary universe with an ordinary state can match either
a closed or an open Euclidean geometry at the cosmological
singularity. This surgery leads inevitably to the instantons
singular at the boundary. The open universe with an exotic state
has the boundary of a finite radius as a turning point of the
classical motion. In the classically forbidden regime, the open
universe is matched not with a closed Euclidean geometry but with
an open Euclidean geometry. The classical equation of motion for
the open Euclidean geometry describes exactly the instanton motion
for the open inflationary universe. It is worthy to note that
there are six more different topologies other than $R^3$ for the
open FRW universe of the Lorentzian geometry, which are foliated
into compact three-manifolds \cite{reboucas}. So this surgery of
the open inflationary universe with the open Euclidean geometry
may not raise any new problem in the Hartle-Hawking wave function
that is defined a sum over different topologies of gravitational
instantons of compact Euclidean manifolds without boundary, which
will be studied elsewhere.

\acknowledgements

The author would like to thank M. J. Rebou\c{c}as for useful
discussions on different topologies of FRW spacetime manifold.
This work was supported by the Korea Research Foundation under
Grant No. 1998-001-D00364.

\end{document}